\documentclass[twocolumn,showpacs,preprintnumbers,amsmath,amssymb]{revtex4}
\usepackage{graphicx}% Include figure files
\usepackage{dcolumn}% Align table columns on decimal point
\usepackage{bm}% bold math

\begin{document}

\preprint{}

\title{Real-Valued Charged Fields\\ and Interpretation of Quantum Mechanics}% Force line breaks with \\

\author{A. M. Akhmeteli}
 \email{akhmeteli@home.domonet.ru}
\affiliation{%
Microinform Training Center\\
38-2-139 Rublevskoe shosse\\
121609 Moscow, Russia}%

 \homepage{http://www.akhmeteli.ru}
%\affiliation{

\date{\today}% It is always \today, today,

\begin{abstract}
Schr\"{o}dinger (Nature, v.169, p.538 (1952)) demonstrated that, contrary to the widespread belief, charged particles may  be described by real fields. Therefore the sets of solutions with real-valued charged fields are considered in the present work for some versions of (non-second-quantized) quantum electrodynamics (for Dirac spinors "real-valued" is understood as "satisfying the Majorana condition"). In some of the versions any solution may be obtained from a solution from those sets by a gauge transform. The solutions from those sets have common features suggesting a natural interpretation along the lines of the Bohm interpretation, but no quantum potentials arise, and it is the electromagnetic field, not the wave function, that plays the role of the guiding field.
\end{abstract}

\maketitle

\section{\label{sec:level1}Introduction}

Using the example of the Klein-Gordon-Maxwell electrodynamics, Schr\"{o}dinger (Ref.~\cite{Schroed}) demonstrated that, contrary to the widespread belief, charged particles may  be described by real fields. This is a part of the rationale for the present work, where the sets of solutions with real-valued electron-positron fields are considered for the Klein-Gordon-Maxwell electrodynamics, the Dirac-Maxwell electrodynamics, and a theory with the Lagrangian of the Dirac-Maxwell electrodynamics subject to the constraint that the axial current vanishes (for Dirac spinors "real-valued" is understood as "satisfying the Majorana condition"). In the first and the third cases any solution may be obtained from a solution from those sets by a gauge transform. In all three cases the solutions from those sets have the following common features: in each point the current is codirectional with the 4-potential of the electromagnetic field; the electromagnetic field satisfies the equations of the Dirac's "new electrodynamics" (Ref.~\cite{Dirac}) with various gauge conditions and displays independent dynamics. These features suggest a natural interpretation along the lines of the Bohm interpretation (Refs.~\cite{BohmHiley,Holland,Goldstein}), but no quantum potentials arise, and it is the electromagnetic field, not the wave function, that plays the role of the guiding field. A "vacuum polarization" version of the Bohm interpretation is considered, where a wave function describes (infinitely) many particles moving along the Bohm trajectories and characterizes the polarization of vacuum. This version seems especially appropriate for real-valued charged fields. The potential implications of this version for the Barut's self-field electrodynamics (Ref.~\cite{Barut}) are considered.
\maketitle

\section{\label{sec:level1}The Dirac's "new electrodynamics"}

This work heavily uses the results of Refs.~\cite{Dirac,Schroed}, so let us summarize some of them here. In Ref.~\cite{Dirac}, Dirac considers the following conditions of stationary action for the free electromagnetic field Lagrangian subject to the constraint $A_\mu A^\mu=k^2$:
\begin{equation}\label{eq:pr1}
\Box A_\mu-A^\nu_{,\nu\mu}=\lambda A_\mu,
\end{equation}
where $A^\mu$ is the potential of the electromagnetic field, and $\lambda$ is a Lagrange multiplier. The constraint represents a nonlinear gauge condition. One can assume that the conserved current in the right-hand side of Eq.~(\ref{eq:pr1}) is created by particles of mass $m$, charge $e$, and momentum (not generalized momentum!) $p^\mu=\zeta A^\mu$, where $\zeta$ is a constant. If these particles move in accordance with the Lorentz equations
\begin{equation}\label{eq:pr2}
\frac{dp^\mu}{d\tau}=\frac{e}{m}F^{\mu\nu}p_\nu,
\end{equation}
where $F^{\mu\nu}=A^{\nu,\mu}-A^{\mu,\nu}$ is the electromagnetic field, and $\tau$ is the proper time of the particle ($(d\tau)^2=dx^\mu dx_\mu$), then
\begin{equation}\label{eq:pr3}
\frac{dp^\mu}{d\tau}=p^{\mu,\nu}\frac{dx_\nu}{d\tau}=\frac{1}{m}p_\nu p^{\mu,\nu}=\frac{\zeta^2}{m}A_\nu A^{\mu,\nu}.
\end{equation}
Due to the constraint, $A_\nu A^{\nu,\mu}=0$, so
\begin{equation}\label{eq:pr4}
A_\nu A^{\mu,\nu}=-A_\nu F^{\mu\nu}=-\frac{1}{\zeta}F^{\mu\nu}p_\nu.
\end{equation}
Therefore, Eqs.~(\ref{eq:pr2},\ref{eq:pr3},\ref{eq:pr4}) are consistent if $\zeta=-e$, and then $p_\mu p^\mu=m^2$ implies $k^2=\frac{m^2}{e^2}$ (so far the discussion is limited to the case $-e A^0=p^0>0$).

Thus, Eq.~(\ref{eq:pr1}) with the gauge condition
\begin{equation}\label{eq:pr5}
A_\mu A^\mu=\frac{m^2}{e^2}
\end{equation}
describes both independent dynamics of electromagnetic field and consistent motion of charged particles in accordance with the Lorentz equations. The words "independent dynamics"  mean the following: if values of the spatial components $A^i$ of the potential ($i=1,2,3$) and their first derivatives with respect to $x^0$, $\dot{A}^i$, are known in the entire space at some moment in time ($x^0=const$), then $A^0$, $\dot{A}^0$ may be eliminated using Eq.~(\ref{eq:pr5}), $\lambda$ may be eliminated using Eq.~(\ref{eq:pr1}) for $\mu=0$ (the equation does not contain second derivatives with respect to $x^0$ for $\mu=0$), and the second derivatives with respect to $x^0$, $\ddot{A}^i$, may be determined from Eq.~(\ref{eq:pr1}) for $\mu=1,2,3$.

\maketitle

\section{\label{sec:level1}Real-valued solutions of the Klein-Gordon-Maxwell electrodynamics}

In his comment on the Dirac's work, Schr\"{o}dinger (Ref.~\cite{Schroed}) considered interacting  scalar charged field $\psi$ and electromagnetic field $F^{\mu\nu}$ with the Lagrangian
\begin{eqnarray}\label{eq:pr6}
\nonumber
-\frac{1}{4}F^{\mu\nu}F_{\mu\nu}+\frac{1}{2}(\psi^*_{,\mu}-ieA_\mu\psi^*)(\psi^{,\mu}+ieA^\mu\psi)-\\
-\frac{1}{2}m^2\psi^*\psi
\end{eqnarray}
and the Klein-Gordon-Maxwell equations of motion
\begin{equation}\label{eq:pr7}
(\partial^\mu+ieA^\mu)(\partial_\mu+ieA_\mu)\psi+m^2\psi=0,
\end{equation}
\begin{equation}\label{eq:pr8}
\Box A_\mu-A^\nu_{,\nu\mu}=j_\mu,
\end{equation}
\begin{equation}\label{eq:pr9}
j_\mu=ie(\psi^*\psi_{,\mu}-\psi^*_{,\mu}\psi)-2e^2 A_\mu\psi^*\psi.
\end{equation}
For each solution $A^\mu$, $\psi$ of these equations there is a physically equivalent (i.e. coinciding with it up to a gauge transform) solution $B^\mu$, $\varphi$, where $\varphi$ is real. For real scalar field the equations of motions may be written in the following form (see also Ref.~\cite{Deumens2}):
\begin{equation}\label{eq:pr10}
\Box\varphi-(e^2 B^\mu B_\mu-m^2)\varphi=0,
\end{equation}
\begin{equation}\label{eq:pr11}
\Box B_\mu-B^\nu_{,\nu\mu}=j_\mu,
\end{equation}
\begin{equation}\label{eq:pr12}
j_\mu=-2e^2 B_\mu\varphi^2.
\end{equation}
Schr\"{o}dinger emphasized two circumstances. Firstly, except for the missing constraint, the equations for the electromagnetic potentials coincide with Eq.~(\ref{eq:pr1}) (if we replace $B_\mu$ with $A_\mu$ and $-2e^2\varphi^2$ with $\lambda$). Secondly, the fact that the scalar field can be made real by a change of gauge, although easy to understand, contradicts the widespread belief about charged fields requiring complex representation. This consideration may be regarded as a part of the rationale for the present work, where real fields are used to describe charged particles.

Obviously, the equations for $B_\mu$ and $\varphi$ are not gauge invariant, as the gauge has already been fixed by the condition that $\varphi$ is real -- unitary gauge (Ref.~\cite{Deumens2},\cite{Itzykson}). It should be noted that these equations may be obtained from the following Lagrangian :
\begin{equation}\label{eq:pr12a}
-\frac{1}{4}F^{\mu\nu}F_{\mu\nu}+\frac{1}{2}e^2 B_\mu B^\mu \phi^2+\frac{1}{2}(\phi_{,\mu}\phi^{,\mu}-m^2\phi^2).
\end{equation}
Actually, it coincides with the Lagrangian of Eq.~(\ref{eq:pr6}) up to the replacement of the complex scalar field by a real one. There is little doubt Lagrangian of Eq.~(\ref{eq:pr12a}) has been considered before, but so far the author has not been able to find the reference.

Rather surprisingly, Eqs.~(\ref{eq:pr10},\ref{eq:pr11},\ref{eq:pr12}) also describe independent dynamics of electromagnetic field in the following sense (assuming $\varphi$ and $B^0$ do not vanish identically): if components $B^\mu$ of the potential and their first derivatives with respect to $x^0$, $\dot{B}^\mu$, are known in the entire space at some moment in time ($x^0=const$), Eqs.~(\ref{eq:pr10},\ref{eq:pr11},\ref{eq:pr12}) yield the values of their second derivatives, $\ddot{B}^\mu$, for the same value of $x^0$, so integration yields $B^\mu$ for any value of $x^0$ (the author is not aware if this has been shown before). Indeed, $\varphi$ may be eliminated using Eq.~(\ref{eq:pr11}) for $\mu=0$, as this equation does not contain $\ddot{B}^\mu$ for this value of $\mu$:
\begin{equation}\label{eq:pr13}
\varphi=\sqrt{(-2e^2 B_0)^{-1}(\Box B_0-B^\nu_{,\nu 0})}.
\end{equation}
Then $\ddot{B}^i$ ($i=1,2,3$) may be determined by substitution of Eqs.~(\ref{eq:pr12},\ref{eq:pr13}) into Eq.~(\ref{eq:pr11}) for $\mu=1,2,3$. Conservation of current implies
\begin{equation}\label{eq:pr14}
0=\partial_\mu(B^\mu \varphi^2)=(\partial_\mu B^\mu)\varphi^2+2 B^\mu\varphi\partial_\mu\varphi,
\end{equation}
or
\begin{equation}\label{eq:pr15}
0=(\partial_\mu B^\mu)\varphi+2 B^\mu\partial_\mu\varphi=(\dot{B^0}+B^i_{,i})\varphi+2B^0\dot{\varphi}+2B^i\varphi_{,i}.
\nonumber
\end{equation}
This equation determines $\dot{\varphi}$, as spatial derivatives of $\varphi$ may be found from Eq.~(\ref{eq:pr13}). Differentiation of this equation yields
\begin{eqnarray}\label{eq:pr16}
\nonumber
0=(\ddot{B}^0+\dot{B}^i_{,i})\varphi+(\dot{B}^0+B^i_{,i})\dot{\varphi}+\\
+2(\dot{B}^0\dot{\varphi}+B^0\ddot{\varphi}+\dot{B}^i\varphi_{,i}+B^i\dot{\varphi}_{,i}).
\end{eqnarray}
After substitution of $\varphi$ from Eq.~(\ref{eq:pr13}), $\dot{\varphi}$ from the previous equation, and $\ddot{\varphi}$ from  Eq.~(\ref{eq:pr10}) into  Eq.~(\ref{eq:pr16}), the latter equation determines $\ddot{B^0}$ as a function of $B^\mu$, $\dot{B}^\mu$ and their spatial derivatives (again, spatial derivatives of $\varphi$ and $\dot{\varphi}$ may be found from the expressions for $\varphi$ and $\dot{\varphi}$ as functions of $B^\mu$ and $\dot{B}^\mu$). Thus, if $B^\mu$ and $\dot{B}^\mu$ are known in the entire space at a certain value of $x^0$, then $\ddot{B}^\mu$ may be calculated for the same $x^0$ and, by integration, in the entire space-time. Therefore, we do have independent dynamics of electromagnetic field, although one cannot choose arbitrary values of $B^\mu$ and $\dot{B}^\mu$ at a certain moment in time as, for example, the argument of the square root in Eq.~(\ref{eq:pr13}) must not be negative. However, there is no need to prove that the set of solutions of the relevant equations is rich enough, as it includes all solutions of the Klein-Gordon-Maxwell equations Eqs.~(\ref{eq:pr7},\ref{eq:pr8},\ref{eq:pr9}) (up to a gauge transform). Obviously, the equations of independent dynamics of electromagnetic field can be written in a covariant form. If $\varphi$ or $B^0$ vanish identically, the dynamics of electromagnetic field is also independent but different.

This result may be relevant to interpretation of quantum mechanics. It allows a natural deterministic interpretation along the lines of the Bohm (de Broglie-Bohm) interpretation (Refs.~\cite{BohmHiley,Holland,Goldstein}). Cutting some corners, one may say that in this interpretation (for one particle) the charged field represents an ensemble of point-like particles guided by the field and moving along the lines of current. For example, for the Klein -Gordon field, the current is defined by Eq.~(\ref{eq:pr9}) or Eq.~(\ref{eq:pr12}). The results of this work suggest that electromagnetic field may be regarded as the guiding field, and in each point the particles move along the potential $B^\mu$. This simplification may make the interpretation more attractive. One may ask if it is possible to get rid of particles altogether and leave just electromagnetic field. The author does not think so, but cannot exclude this possibility for reasons outlined later in this work along with a somewhat different interpretation.

It should be mentioned that there is some controversy about the Bohm interpretation of the Klein-Gordon field, in particular, because current may be spacelike for this field. For example, in Refs.~\cite{Nik1,Nik2} it is contended that these difficulties do not lead to inconsistencies; a different definition of particle trajectories is given in Refs.~\cite{Dewd1,Dewd2} (see, however, Ref.~\cite{Tum2}); there is also an opinion that bosons are fields, and they have no particle trajectories (Ref.~\cite{Holland}).  This author focuses, however, on electrodynamics and does not consider any massive bosons, so the Klein-Gordon equations are regarded just as a reasonably decent approximation for electrons. Therefore, the inevitable next step would be to replace the Klein-Gordon field by the Dirac field.
\maketitle

\section{\label{sec:level1}Majorana solutions of the Dirac-Maxwell electrodynamics}

The Schr\"{o}dinger's remark on the possibility of description of charged particles with real fields suggests that charged particles of spin 1/2 may be described by Majorana spinors (actually, Majorana developed his theory (Ref.~\cite{Majorana}) for electrons), as the Majorana condition is an analog of the reality condition and coincides with the latter in the Majorana representation of $\gamma$-matrices (Ref.~\cite{Itzykson}). So in this work the Dirac-Maxwell electrodynamics  is considered:
\begin{equation}\label{eq:pr17}
(i\hat{\partial}-e\hat{A}-m)\Psi=0,
\end{equation}
\begin{equation}\label{eq:pr18}
\Box A_\mu-A^\nu_{,\nu\mu}=j_\mu,
\end{equation}
\begin{equation}\label{eq:pr19}
j_\mu=e\bar{\Psi}\gamma_\mu\Psi,
\end{equation}
and the subset of its solutions is defined by the condition that $\Psi$ is a Majorana spinor. It was shown in Ref.~\cite{Buchm} that this subset is not trivial. Two types of spinors were considered (Refs.~\cite{Akhm1},\cite{Akhm2},\cite{Akhm3},\cite{Akhm4}; Ref.~\cite{Akhm2}, its English translation, Ref.~\cite{Akhm3} and Ref.~\cite{Akhm4} may be found at the author's web site): $c$-type spinors (components of the spinor $\Psi$ are $c$-numbers) and $a$-type spinors (components of the spinor $\Psi$ are anticommuting elements of a Grassman algebra), but the results are illustrated here using c-spinors, so the current for a Majorana spinor only vanishes if the spinor equals zero.

Applying charge conjugation to the Dirac equation (Eq.~(\ref{eq:pr17})) and using the Majorana condition, we obtain $(i\hat{\partial}-m)\Psi=0$ and $\hat{A}\Psi=0$. The latter equation implies $A_\mu A^\mu=0$, if $\Psi\neq 0$ (Ref.~\cite{Buchm}); furthermore, if the vector $A^\mu$ is not zero, the equation also implies that there exists such $\lambda$ that $j^\mu =\lambda A^\mu$ (Ref.~\cite{Buchm}). It should be noted that $j_\mu j^\mu=0$ for Majorana spinors (Ref.~\cite{Deumens1}). We also obtain
\begin{eqnarray}\label{eq:pr20}
0=(\hat{\partial}\hat{A}+\hat{A}\hat{\partial})\Psi=2A^\mu\partial_\mu\Psi+A_{\nu,\mu}\gamma^\mu\gamma^\nu\Psi.
\end{eqnarray}
This equation may be regarded as a system of ordinary (not partial!) differential equations on a curve that is tangential in all its points $x$ to the vector $A^\mu(x)$.

We may conclude that the equations of the Maxwell-Dirac electrodynamics (Eqs.~(\ref{eq:pr17},\ref{eq:pr18},\ref{eq:pr19})) for Majorana $c$-type spinors are equivalent (if $\Psi$ and $A$ do not vanish identically) to the following system:
\begin{equation}\label{eq:pr21}
(i\hat{\partial}-m)\Psi=0,
\end{equation}
\begin{equation}\label{eq:pr22}
A_\mu A^\mu=0,
\end{equation}
\begin{equation}\label{eq:pr23}
\lambda A^\mu=j^\mu =e\bar{\Psi}\gamma^\mu\Psi,
\end{equation}
\begin{equation}\label{eq:pr24}
\Box A_\mu-A^\nu_{,\nu\mu}=\lambda A_\mu.
\end{equation}
Eqs.~(\ref{eq:pr24},\ref{eq:pr22}) coincide with Eqs.~(\ref{eq:pr1},\ref{eq:pr5}) of the Dirac's "new electrodynamics" (Ref.~\cite{Dirac}) up to a constant in the right-hand side of Eq.~(\ref{eq:pr22}). A solution of Eq.~(\ref{eq:pr24}) satisfies the condition of stationary action of the free electromagnetic field subject to the constraint Eq.~(\ref{eq:pr22}) (cf. Ref.~\cite{Dirac}). Therefore, Eqs.~(\ref{eq:pr24},\ref{eq:pr22}) describe independent evolution of the electromagnetic field ($\lambda=\lambda(x)$ is a Lagrangian multiplier). Eq.~(\ref{eq:pr22}) may be regarded as a nonlinear gauge condition.
The subset of the Dirac-Maxwell electrodynamics allows a natural deterministic interpretation along the lines of the Bohm interpretation. The difference is that the quantum potential(s) is/are replaced by the ordinary potential of electromagnetic field (or, in a version of the Bohm interpretation where a quantum potential does not play a prominent role (Ref.~\cite{Goldstein} and references there), electromagnetic field replaces wave function as a guiding field). An electron may be regarded as a point-like particle with properties that are determined solely by the value of the spinor $\Psi$ in the point of space-time. Possible trajectories are the curves that are tangential to the vector $A^\mu(x)$ in every their point $x$. Eq.~(\ref{eq:pr22}) implies that the magnitude of the instantaneous velocity is always equal to the velocity of light. This is consistent with the uncertainty principle and the notion of zitterbewegung and allows smaller mean velocities (the role of zitterbewegung for interpretation of quantum mechanics was discussed in a great number of works. See, e.g., Refs.~\cite{Hestenes},\cite{Perkins},\cite{Corben}).

It seems that there may exist a somewhat different interpretation of real charged fields: the one-particle $\Psi$-function may describe a large (infinite?) number of particles moving along the above-mentioned trajectories. The total charge, calculated as an integral of charge density over the infinite 3-volume, may still equal the charge of electron. So the individual particles may be either electrons or positrons, but all together they may be regarded as one electron, as the total charge is conserved (if an electron is then removed, for example, as a result of a measurement, and the total energy of what is left is not very high, so it is difficult to speak about presence of pairs, then the remaining field will look very much like electronic vacuum, maybe with some electromagnetic  field). This seems to be compatible with the notions of polarization of vacuum and path integral. So $\Psi(x)$ may be a measure of polarization of vacuum in the point $x$ (and this may explain the fact that it determines the density of probability of finding a particle in this point), and spreading of wave packets should not create problems. This interpretation also seems to give a clearer picture of the two-slit interference. The author is not sure if such an interpretation has been proposed for ordinary complex charged fields, but it seems especially appropriate for real charged fields.

As the system of Eqs.~(\ref{eq:pr21},\ref{eq:pr22},\ref{eq:pr23},\ref{eq:pr24}) is overdetermined, it is necessary to find an equivalent involutive system (Ref.~\cite{Pommaret}) to find out if the set of its solution is rich enough. Unfortunately, although there are finite algorithms solving this task, the actual calculations may be intractable. Recently some methods have been developed for systems with high degree of symmetry (Ref.~\cite{Mansfield}) that may help overcome this difficulty. However, so far this problem has not been solved. Therefore it may be advisable to make a step in a different direction. So far we have only considered solutions of well-established theories -- the Klein-Gordon-Maxwell and Dirac-Maxwell electrodynamics, so in this respect we have been on firm ground, no matter how controversial their interpretation may be. Now let us consider the standard Lagrangian of the Dirac-Maxwell electrodynamics and impose the constraint $\bar{\Psi}\gamma^5\gamma^\mu\Psi=0$ (the axial current vanishes). A similar approach to imposition of the Majorana condition was used in Ref.~\cite{Lochak}, but the specific procedure there raises some doubts as the constraints of that work make no contribution to the equations of motion. In our case the equations of motion are as follows:
\begin{equation}\label{eq:pr25}
(i\hat{\partial}-e\hat{A}+\gamma^5\hat{D}-m)\Psi=0,
\end{equation}
\begin{equation}\label{eq:pr26}
\Box A_\mu-A^\nu_{,\nu\mu}=j_\mu,
\end{equation}
\begin{equation}\label{eq:pr27}
j_\mu=e\bar{\Psi}\gamma_\mu\Psi,
\end{equation}
\begin{equation}\label{eq:pr28}
\bar{\Psi}\gamma^5\gamma^\mu\Psi=0,
\end{equation}
where $\hat{D}=D_\mu\gamma^\mu$, and $D_\mu$ are the Lagrangian multipliers. Every solution of this system is physically equivalent to a Majorana solution related to it via a gauge transform: Eq.~(\ref{eq:pr28}) implies that the spinor $\Psi$ may be represented in the form $\Psi=\exp(i\theta)\Phi$, where $\theta=\theta(x)$ is real, and $\Phi$ is a spinor satisfying the Majorana condition. Substituting this in Eqs.~(\ref{eq:pr25},\ref{eq:pr26},\ref{eq:pr27}), we obtain equations for Majorana spinors:
\begin{equation}\label{eq:pr29}
(i\hat{\partial}-e\hat{B}+\gamma^5\hat{D}-m)\Phi=0,
\end{equation}
\begin{equation}\label{eq:pr30}
\Box B_\mu-B^\nu_{,\nu\mu}=j_\mu,
\end{equation}
\begin{equation}\label{eq:pr31}
j_\mu=e\bar{\Phi}\gamma_\mu\Phi,
\end{equation}
where $e B_\mu=e A_\mu+\theta_{,\mu}$. Treating Eq.~(\ref{eq:pr29}) in the same way as the Dirac equation (Eq.~(\ref{eq:pr17})), we obtain:
\begin{equation}\label{eq:pr32}
(i\hat{\partial}+\gamma^5\hat{D}-m)\Phi=0,
\end{equation}
\begin{equation}\label{eq:pr33}
\hat{B}\Phi=0.
\end{equation}
Again, Eq.~(\ref{eq:pr33}) implies $B_\mu B^\mu=0$, if $\Phi\neq 0$; if the vector $B^\mu$ is not zero, the equation also implies that there exists such $\lambda$ that $j^\mu =\lambda B^\mu$. Therefore, we obtain the following system of equations with Majorana spinors:
\begin{equation}\label{eq:pr34}
(i\hat{\partial}+\gamma^5\hat{D}-m)\Phi=0,
\end{equation}
\begin{equation}\label{eq:pr35}
B_\mu B^\mu=0,
\end{equation}
\begin{equation}\label{eq:pr36}
\lambda B^\mu=j^\mu =e\bar{\Phi}\gamma^\mu\Phi,
\end{equation}
\begin{equation}\label{eq:pr37}
\Box B_\mu-B^\nu_{,\nu\mu}=\lambda B_\mu.
\end{equation}
Eq.~(\ref{eq:pr34}) is linear in $D_\mu$, so it is easy to eliminate $D_\mu$ from it. Again, Eqs.~(\ref{eq:pr35},\ref{eq:pr37}) describe independent evolution of the electromagnetic field, and Eq.~(\ref{eq:pr36}) allows one to determine the trajectories in the Bohm interpretation from the potential of the electromagnetic field. It remains to be seen whether Eqs.~(\ref{eq:pr34},\ref{eq:pr35},\ref{eq:pr36},\ref{eq:pr37}) are compatible with experimental data or they may only be used as an interesting toy model for interpretation of quantum mechanics.

It should be noted that second-quantized theories are not considered in this work. Transition to several particles is known to introduce complications in interpretation of quantum theory. So this is definitely a drawback of this work (and certainly not the only one). However, this is all the author can offer at the moment. As a partial justification of his approach, he may say that interpretation of quantum mechanics may be discussed using theories of various levels of sophistication, and the theories used in this work are far from being the most primitive. On the other hand, there is also a possibility to refer to the Barut's self-field electrodynamics (Ref.~\cite{Barut}. See also his ICTP preprints, e.g. Ref.~\cite{Barut2}). Barut contends that calculations in the non-second-quantized Dirac-Maxwell electrodynamics reproduce the famous results of the standard (second-quantized) quantum electrodynamics, e.g. for the Lamb shift, with great accuracy. However, as far as this author understands, Barut put his hopes on finding soliton solutions of the Dirac-Maxwell electrodynamics for description of free electrons. This author has his doubts, but some soliton solutions have been found (Ref.~\cite{Buchm},\cite{Georgiev},\cite{Radford}). However, the "vacuum polarization" version of the Bohm interpretation outlined above may remove the need for soliton solutions in the self-field electrodynamics.
\maketitle

\section{\label{sec:level1}Conclusion}
Some versions of (non-second-quantized) quantum electrodynamics with real-valued charged fields have been considered. All of them have common features suggesting a natural interpretation along the lines of the Bohm interpretation, but no quantum potentials arise, and it is the electromagnetic field, not the wave function, that plays the role of the guiding field.

 \bibliography{maj1}% Produces the bibliography via BibTeX.

\end{document}